\documentclass[twocolumn]{article}
\usepackage{layout}
\usepackage{amsfonts}
\usepackage{graphicx}
\usepackage{url}
\usepackage{hyperref}
 \newcommand{\figura}[4]{ \begin{figure}[htb] \centering \resizebox{#1}{!}{\includegraphics[#4]{#2}}\\
\caption{\label{#2} #3} \end{figure}}
\newcommand{\beq}[1]{\begin{equation}\label{eq:#1}}
\newcommand{\eeq}{\end{equation}}
\newcommand{\refig}[1]{(Figure: \ref{#1})}
\usepackage{amssymb}
\usepackage{lineno}
\newcommand{\refsec}[1]{\textbf{(Sec: \ref{sec:#1}})}
\begin{document}
\title{Size distribution and waiting times for the 
  avalanches of the Cell Network Model of Fracture}
\author{Gabriel Villalobos$^{a}$\footnote{Corresponding author, gabrielvc@gmail.com}, Ferenc Kun$^{c}$, Dorian L. Linero$^{b}$, Jos\'e D. Mu{\~n}oz$^{a}$\\*\footnotesize{$^{a}$Simulation of Physical Systems Group, CeiBA-Complejidad, Department of Physics,}\\* \footnotesize{Universidad Nacional de Colombia, Crr 30 \# 45-03, Ed.~404, Of.~348, Bogota D.C., Colombia.,}\\*\footnotesize{$^{b}$Analysis, Design and Materials Group, Department of Civil and Environmental Engineering,} \\*\footnotesize{ Universidad Nacional de Colombia, Crr 30 \# 45-03, Ed.~404, Of.~348, Bogota D.C., Colombia.,}\\*
\footnotesize{$^{c}$Department of Theoretical Physics. University of Debrecen, H-4010 Debrecen, P.O.Box 5, Hungary.}}
\maketitle

\begin{abstract}
The Cell Network Model is a fracture model recently introduced that resembles the microscopical structure and drying process of the parenchymatous tissue of the Bamboo \emph{Guadua
  angustifolia}. The model exhibits a power-law distribution of avalanche sizes, with exponent $-3.0$ when the breaking thresholds are randomly distributed with uniform probability density. Hereby we show that the same exponent also holds when the breaking thresholds obey a broad set of Weibull distributions, and that the humidity decrements between successive avalanches (the equivalent to waiting times for this model) follow in all cases an exponential distribution. Moreover, the fraction of remaining junctures shows an exponential decay in time. In addition, introducing partial breakings and cumulative damages induces a crossover behavior
between two power-laws in the avalanche size histograms. This results support the idea that the Cell Network Model may be in the same universality class as the Random Fuse Model.

Statistical models of fracture Finite Element Method Computational mechanics of solids.

\end{abstract}




\section{Introduction}
\label{sec:intro}

The Cell Network Model of Fracture (CNMF \cite{VILLALOBOS10}), is a
two dimensional statistical model of fracture \cite{alava-2006-55}
inspired in the stress field caused by drying of the bamboo
\emph{Guadua angustifolia} \cite{Montoya06,Takeuchi2008}. At the
parenchymatous tissue level, bamboos shrink during drying, causing the
detaching of neighboring cells and the appearance of fractures. The
CNMF models this tissue as an hexagonal array of cell elements (each
of them made of six beams), fixed by angular springs and joined by
brittle springs called the junctures. (Figure: \ref{bamboo.pdf}).
Shrinking forces -due to drying- acting along the elements distort the
structure and cause breaking avalanches of the junctures among cells.

\figura{8cm}{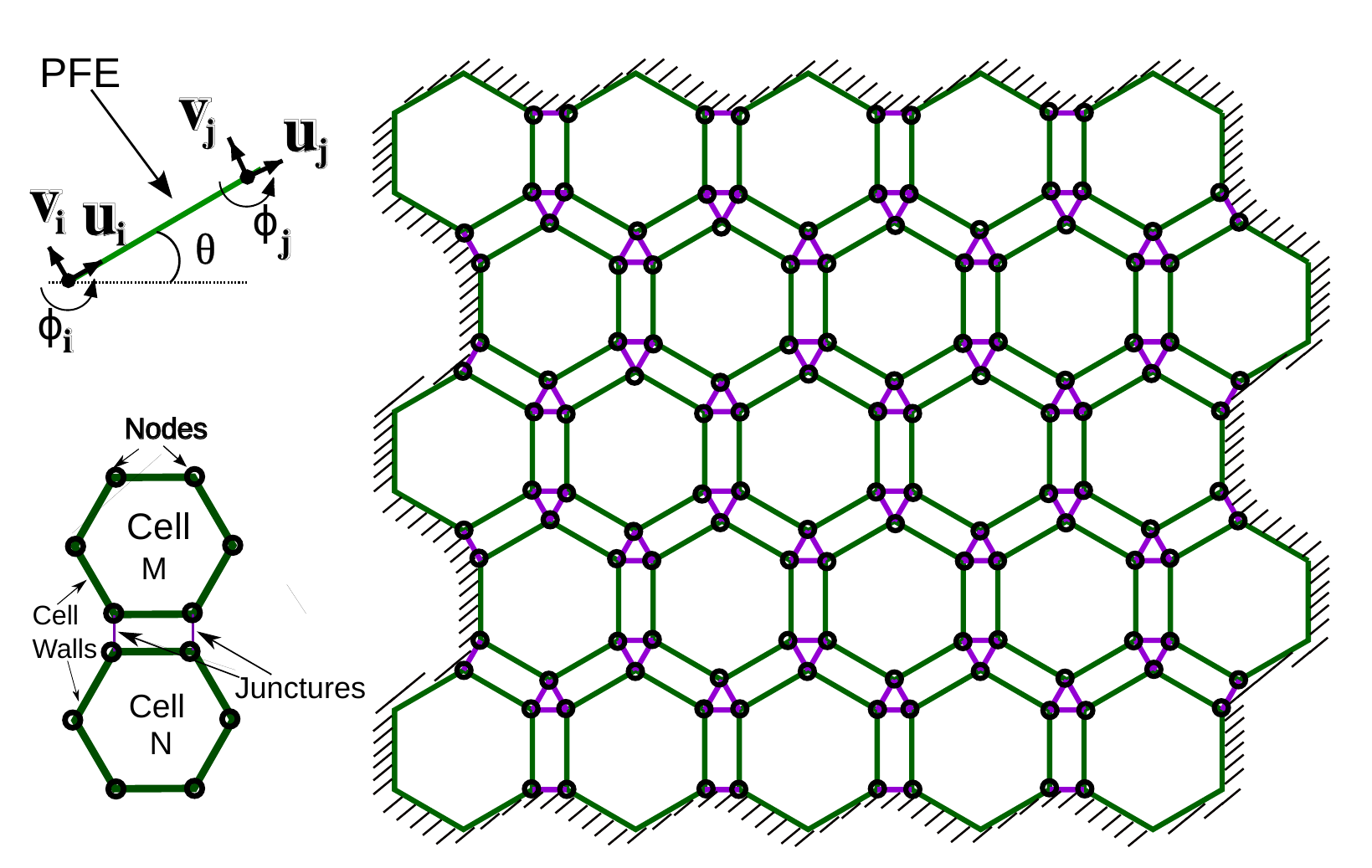}{Cell Network Model of Fracture
  (CNMF). \emph{Upper left}, The plane frame element spanning between
  the nodes $i$ and $j$, oriented by an angle $\theta$. Each node has
  two translational and one rotational degrees of freedom. \emph{Lower
    left}. Two contiguous Cells. \emph{Right} Structure of the
  CNMF. The hexagons represent the cells and the junctures are
  arranged into triangles.  Hashing denote fixed boundary
  conditions. From \cite{VILLALOBOS10} Not at scale.}{}

Interestingly enough, when an homogeneous distribution of breaking
thresholds of junctures and fixed Young modulii are used, the
histogram of avalanche sizes of the CNMF shows power law behavior with
an exponent of $-2.93(8)$ (\cite{VILLALOBOS10}). This is by all means
equal to avalanche size distribution of the random fuse model, which
shows a power law with exponent of $-3$
\cite{PhysRevE.74.016122,Hansen94}. The analytical solution of both
models has been elusive.

One dimensional fiber bundle systems have provided models for the
thoroughly study of critical phenomena. Universality, the effect of
damage and the critical exponents have been found analytically (see
\cite{Kun2005,PradhanHansenEtAlFailure09} and references therein.).
The logical extension of those models to 2D, the random fuse model,
has allowed to investigate the fracture properties of biological materials, 
as in the case of brittle nacre, 
(\cite{PhysRevE.72.041919}), describing 
the toughness of the material by its microscopical architecture.  Beam models
similar to the CNMF has also been used to model fracture in concrete
\cite{PhysRevE.75.066109}.

In the present paper, to characterize the path to global failure, we
study both the distribution of humidity decrement among between
consecutive avalanches (the analogue of waiting times for this model),
as well as the fraction of intact fibers as function of the humidity
decrement. This last quantity behaves as an order parameter for the
system (see \refsec{model}).  Moreover, the universality of the CNMF
is explored numerically by characterizing the histogram of avalanche
sizes for two cases: homogeneous distributions of the juncture
breaking thresholds with different widths and Weibull distributions of
different shapes, \refsec{disorder}.  Furthermore, in section
\refsec{damage} a damage function is introduced as follows: Each time
a stress threshold is reached, the stiffness is reduced by a constant
factor, until the fiber completely breaks. The main results and
comments are summarized in Sec \refsec{concludingremarks}.

\section{Model}
\label{sec:model}

The CNMF is a 2D statistical model of fracture that resembles the
parenchymatous tissue of the bamboo {\it Guadua augustifolia}. It is
composed by two kinds of structures: cell walls and junctures among
cells. Six cell walls arrange themselves to form hexagons, thanks to a
angular springs associated with the rotational degree of freedom. The
cells are arranged like a honeycomb. The junctures are arranged in sets
of three at the common corners of the cells, modeling the silica
deposits that glue cells together. (Figure: \ref{bamboo.pdf}) Each
kind has a given fixed Young modulus for all its elements. Junctures
are allowed to break, as a result of the brittle behavior of the
silica, while cell walls are not.

As boundary conditions, all the border nodes are set fixed.  The
deformation of the system comes from shrinking forces acting on every
cell wall and proportional to a global humidity loss parameter $\Delta
h$. By means of a Finite Element Method, the resulting forces and
deformations of all the elements are calculated.

The evolution of the system has three stages. \emph{Linear Elastic
  Shrinking}: local shrinking forces due to humidity losses are
applied to the cell walls. The differences between the local strain at
the juncture and their individual thresholds are calculated. This step
continues until at least one fiber would suffer an strain surpassing
its threshold. \emph{ Drying induced Breaking}: By means of a zero
finding algorithm, the exact humidity loss causing the first breaking
is found. The broken element is removed from the structure, changing
the stiffness matrix in accordance. \emph{Nonlinear avalanche}: The
breaking of an element calls for a force redistribution over the whole
structure. This redistribution may cause an avalanche of breaks. When
the avalanche ends, the procedure re-stars from the first stage.

\subsection{Distributions of humidity decrement between consecutive avalanches}

The distribution of humidity decrements between successive avalanches
provides a description of the path to the global failure of the
system. This is the analog to the waiting times between avalanches of
other models of fracture.  Fig. \refig{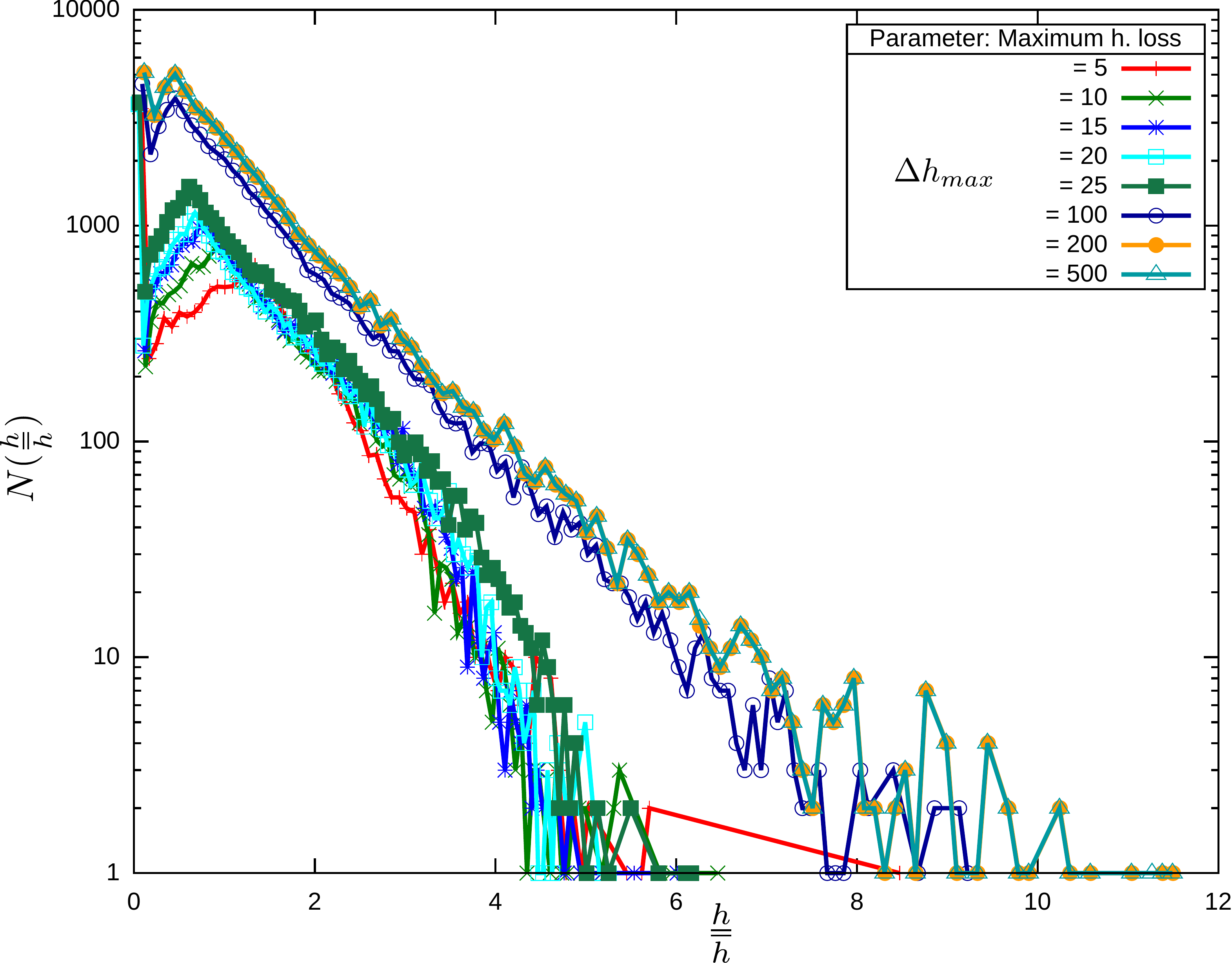} shows the
normalized histograms of humidity decrements for several values of the
maximum humidity change allowed $\Delta h$. When the breaking process
is driven until the end, the histograms can be fitted to an
exponential, with a fitted humidity constant of $13.5(2)$, which is
and indication of lack of correlation between successive avalanches.

\figura{8.5cm}{largsmall.pdf}{Humidity decrement between consecutive
  avalanches histogram. The parameter that defines a curve
  is the maximum humidity change ($\Delta h$, proportional to Cell
  Wall strain, shown). The classes (horizontal axis) are the
  (normalized by the mean waiting humidity).}{}{}

\subsection{Fraction of remaining junctures}


Let us consider the one-dimensional global load sharing fiber bundle model
\cite{PradhanHansenEtAlFailure09}. Given $U(\sigma)$ as the fraction of 
remaining fibers at a given stress and $\sigma_c$ the critical stress causing 
the global breaking of the system. Thus, $U^*(\sigma) - U^*(\sigma_c)$ behaves as an order
parameter.  For the CNMF with homogeneous
breaking thresholds at the juncture elements  we obtain an exponential
relaxation in the number of remaining junctures (\refig{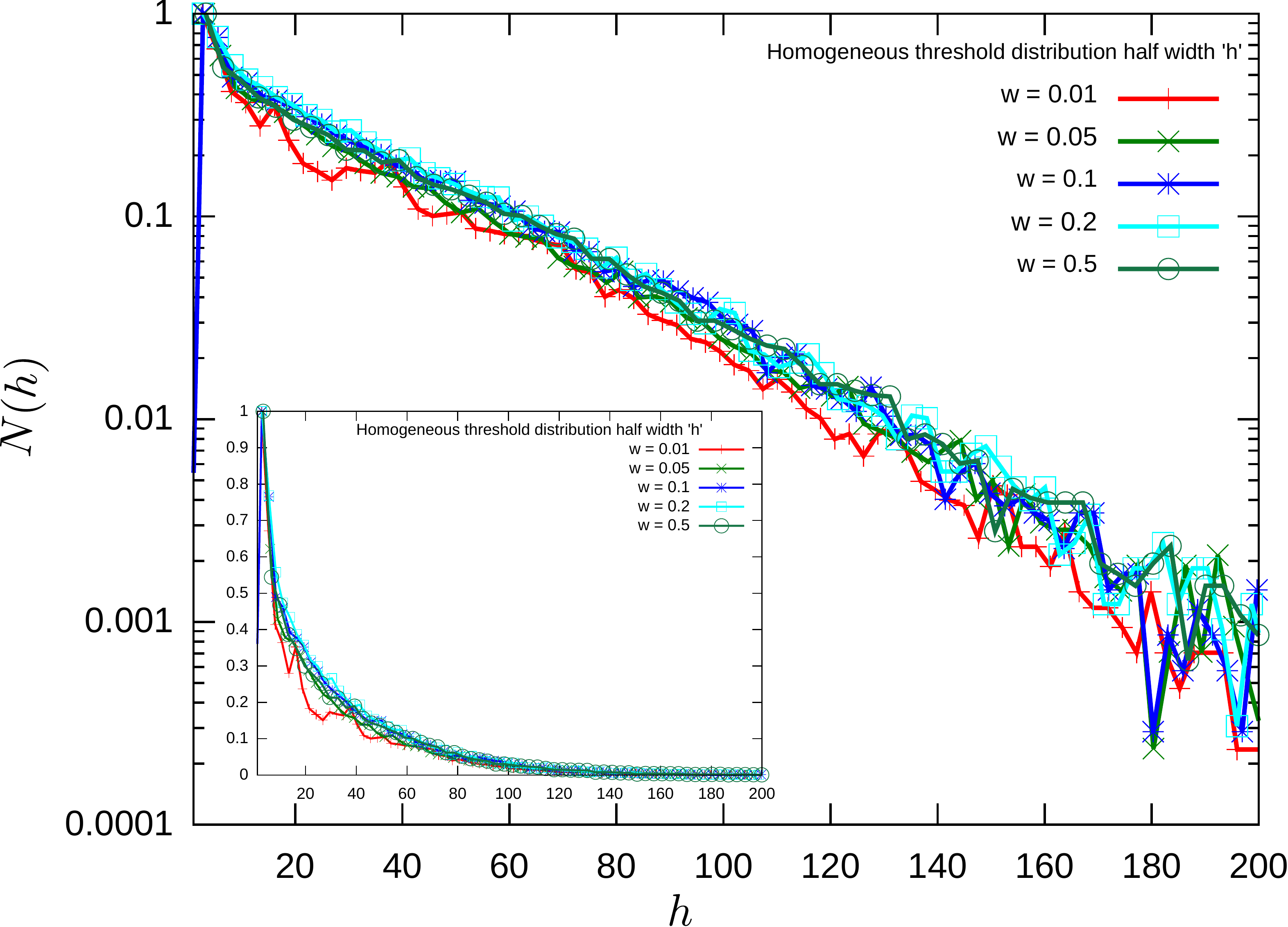}).  

\figura{8.5cm}{varWidthLogY.pdf}{Number of intact fibers as function
  of the humidity change for several homogeneous distribution of the
  breaking threshold with different widths (semi log). \emph{Inset}, linear
  axis. }{}{}

This exponential behavior is independent of the system size. In
\refig{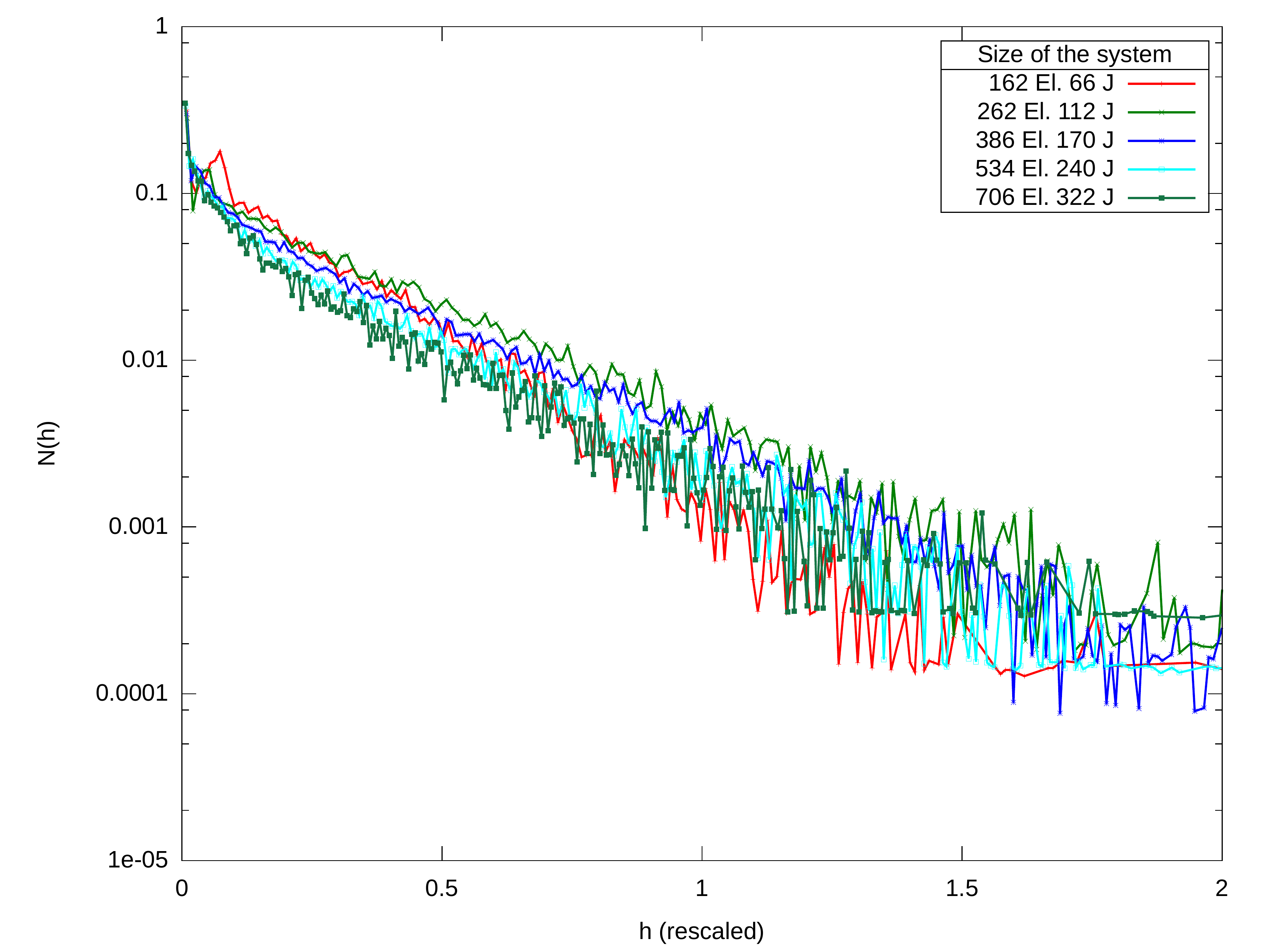}, the fraction of the population of intact
fibers as function of humidity is shown for different sizes of the
system. The horizontal axis was rescaled by means of a linear fit on
the semi-log data. All histograms show an exponential decay. 

\figura{8.5cm}{OrderParScaling.pdf}{Scaled fraction of 
  intact fibers as function of the humidity change (in units of max
  humidity difference) for different system sizes. }{}{}

\section{Universality}
\label{sec:disorder}
Fiber bundle models are universal in the sense that the breaking of the elements follows power law distribution of avalanche sizes irrespective of several system characteristics. For the CNMF we studied the distribution of avalanche sizes when the thresholds are generated either from flat distributions of several widths or from Weibull distributions with several characteristic parameters.

Flat distributions of the breaking thresholds, all centered at $0.35 EA$ but with different widths (spanning on two orders of magnitude), show the same power-law distribution of avalanche sizes, with slopes around $-3$ (Fig.~\refig{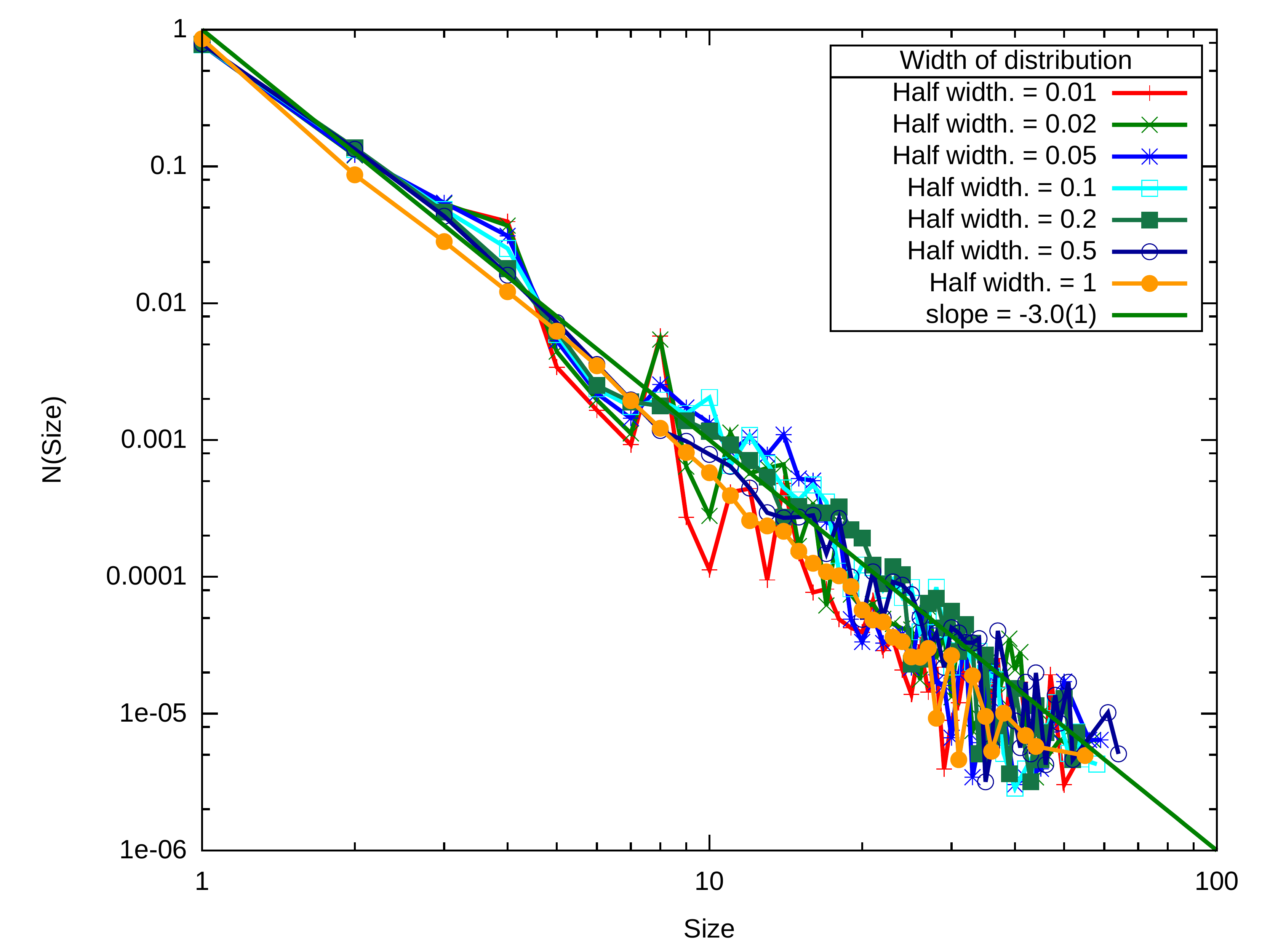}). The data shows that narrower distributions show larger fluctuations around this power law than wider ones.

\figura{8.5cm}{ChangeSize.pdf}{Histogram of avalanche sizes for
  several widths of the homogeneous threshold distribution centered at
  $0.35 EA$}{}{}

The probability density function for the Weibull distribution function
\cite{Weibull51} is given by: \beq{Weibull}
f(x;\lambda,k) = \left\{ \begin{array}{cc} \frac{k}{\lambda}
  \left(\frac{k}{\lambda}\right)^{k-1}e^{-(x/\lambda)^k} & x\geq0\\ 0
  & x<0
\end{array}\right. ,
\eeq
\noindent where $k$ and $\lambda$ are free parameters. It is commonly
used to describe the breaking thresholds of fibers by fixing
$\lambda$$=$$1$ and changing $k$, which is the main parameter
controlling the distribution shape. When we use this distribution for
the breaking thresholds at the junctures, the histogram of avalanche
sizes shows also a power law behavior, with an exponent close to -2.9
for small values of k. For larger values the exponential cutoff is
more pronounced. \refig{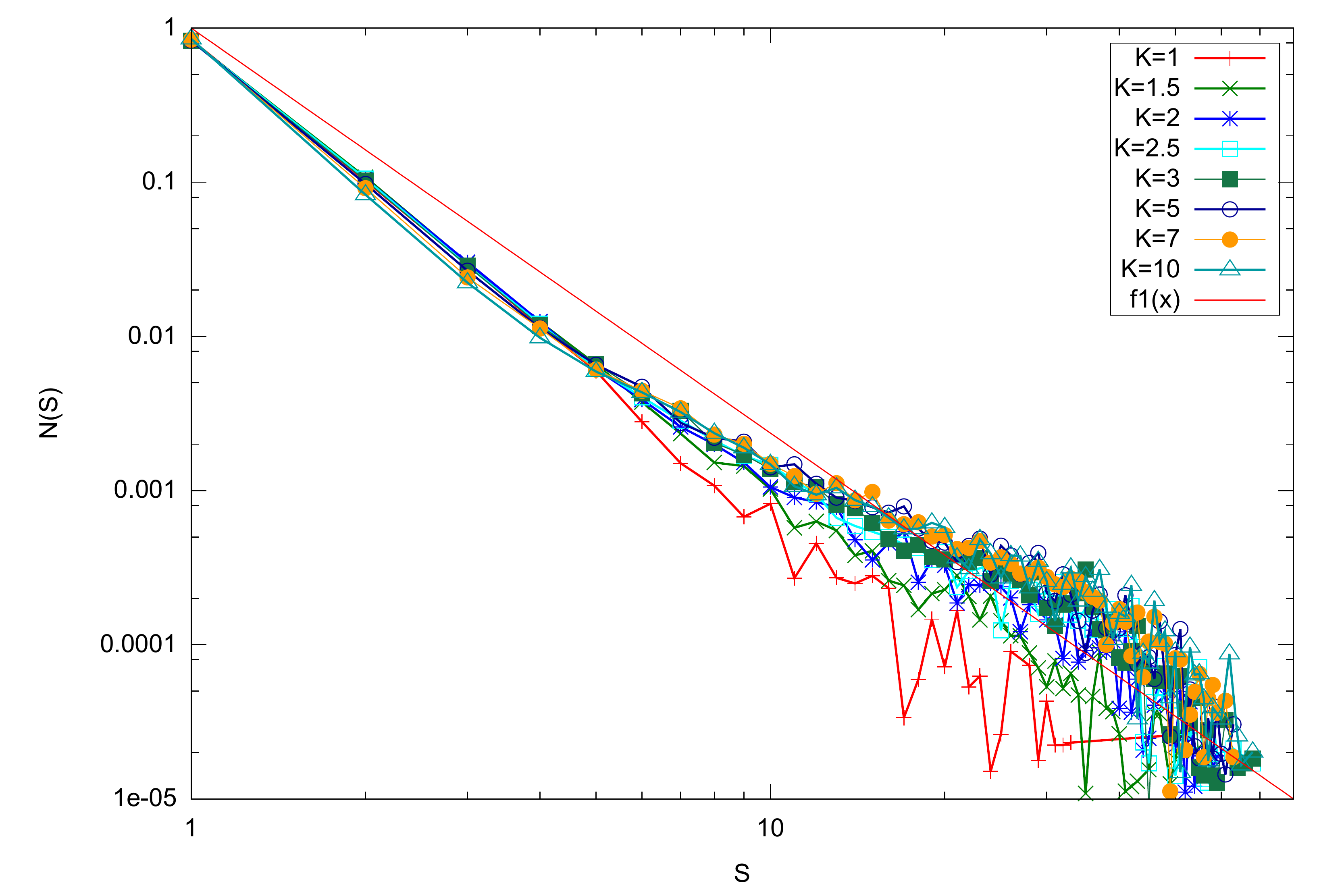}. 

\figura{9.cm}{Weibull1-10.pdf}{Histogram of avalanche sizes for
  several Weibull distributions of the breaking thresholds, with
  values of the shape parameters $k$ between $1$ and $10$. The
  straight line corresponds to an slope of $-2.9$.
}{}{}


The humidity decrements between consecutive avalanches (that is, the
waiting times) distributes like an exponential, also when the
thresholds follow a Weibull distribution (\refig{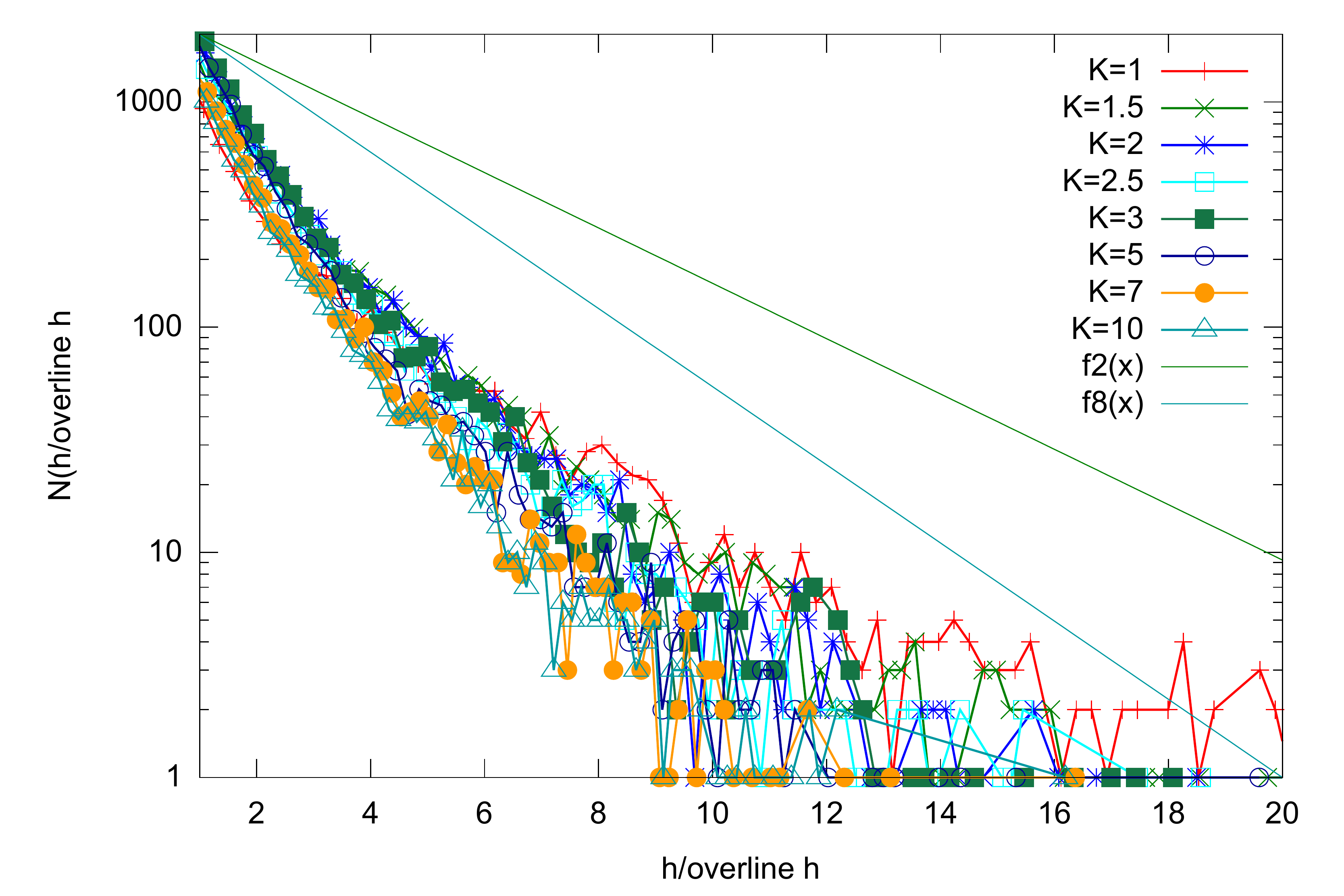}).The
characteristic time for $k$$=$$5$ is 16.1(8), very close to the one
we gathered for flat distributions of the braking thresholds.

\figura{8.5cm}{WTWeibullC.pdf}{Histogram of humidity increments
  between successive avalanches for several Weibull distributions of
  the breaking thresholds, with $k$ between $1$ and $10$. The blue
  line corresponds to a fit of the series of shape $k=10$, with slope
  $-2.1$. The green line to that for $k=3$, with slope $-2.4
$
}{}{}


\section{Damage}
\label{sec:damage}
In order to introduce a degradation for the juncture elements we
reduce the Young modulus of the juncture elements by a damage factor
$0<a<1$ each time the juncture fails. When the element has suffered a
maximum number of failures $k_{max}$, it is assumed to be broken and
is removed from the structure.

\figura{8.5cm}{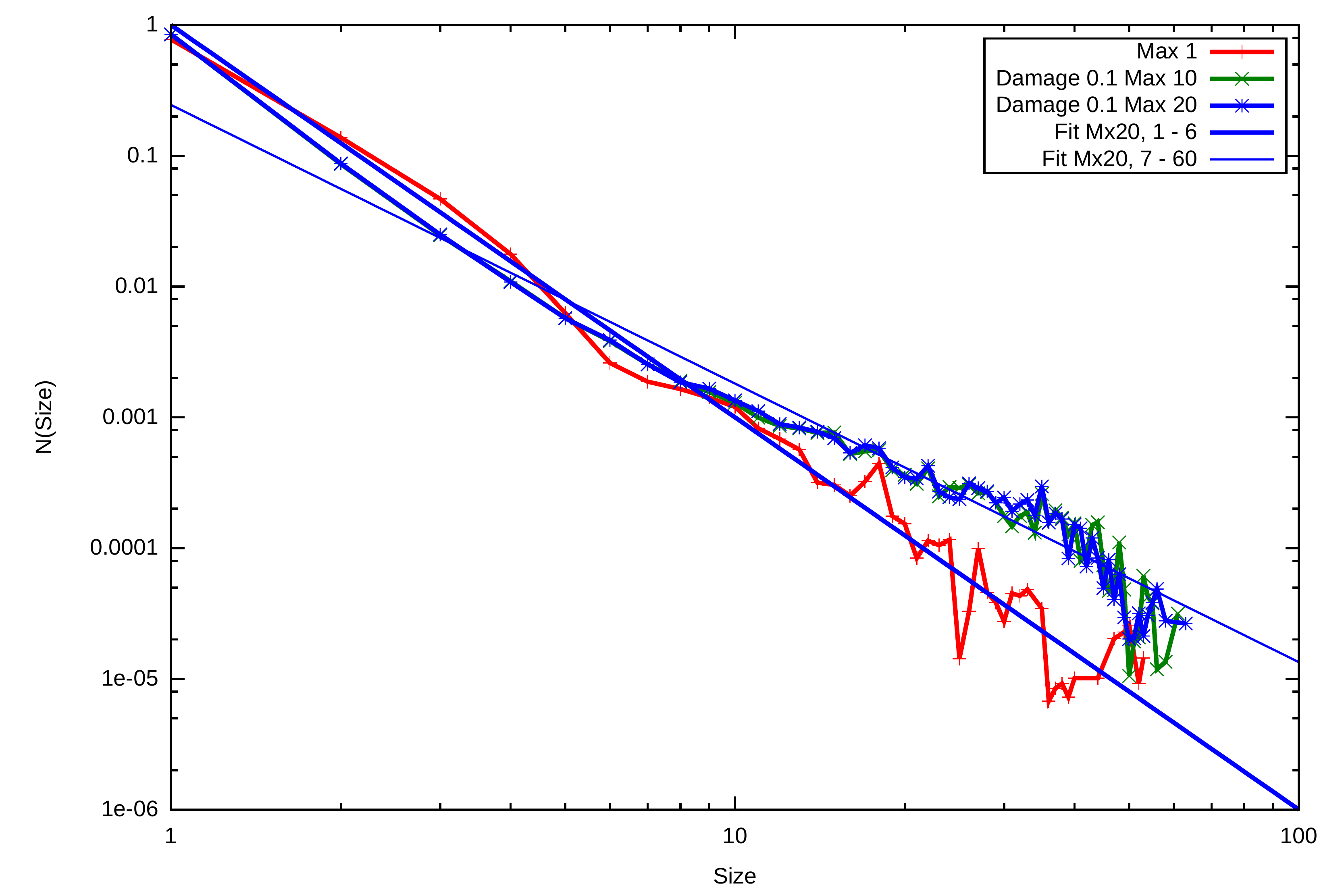}{Histogram of avalanche sizes for damage
  parameter $a$$=$$0.1$ and several maximal numbers of failures, $k_{max}$. The bolder line (slope $-2.98(7)$ fits for avalanche sizes between 1 and 7, while the thinner line (slope $-2.1(1)$) fits for avalanche sizes between 7 and 60.}{}{} 

When a small degradation is introduced ($a$$=$$0.1$), the power law distribution of avalanche sizes seems to exhibit a crossover from an exponent $-3$ to an exponent $-2$ at sizes around 8 (Fig. \refig{CHMxD1.pdf}). This may indicate that the remaining elasticity helps to sustain the structure. However, more statistics and larger system sizes are required to clarify this point.  Even smaller values of $a$ (not shown) cause the maximum humidity loss (and therefore the forces on the elements) to be much larger, creating numerical instabilities that end into poor statistics.

\section{Conclusions}
\label{sec:concludingremarks}
The numerical evidence of this work indicates that the avalanche sizes
for the Cell Network Model of Fracture distribute as a power law with exponent $-3.0$, for any broad distribution of the braking thresholds, either if they are flat or Weibull distributed.

The distribution of waiting times show an exponential decay for all
 system sizes evaluated and all tested disorder
distributions of breaking thresholds. Even the characteristic times are similar for all of them. In our opinion, this is related to the fact that the
most common failure mode for the system is the softening of the sample by
displacements that would violate the boundary conditions.

\textbf{Acknowledgments:} We thank
\href{http://www.colciencias.gov.co}{\emph{COLCIENCIAS}}
(``Convocatoria Doctorados Nacionales 2008''),
\href{http://www.ceiba.org.co}{\emph{Centro de Estudios
    Interdisciplinarios B{\'a}sicos y Aplicados en Complejidad- CeiBA
    - Complejidad}} and
\href{http://www.unal.edu.co}{\emph{Universidad Nacional de Colombia}}
for financial support. We also thank Professor Jorge A. Montoya and
Professor Caori P. Takeuchi for enlightening discussions in the field
of Guadua drying.











\bibliographystyle{elsarticle-num}
\bibliography{VILLALOBOS-SMF-CCP}

\end{document}